\documentclass{article}

\usepackage[english]{babel}

\usepackage[letterpaper,top=2cm,bottom=2cm,left=3cm,right=3cm,marginparwidth=1.75cm]{geometry}

\usepackage{amsmath}
\usepackage{graphicx}
\usepackage[colorlinks=true, allcolors=blue]{hyperref}
\usepackage{rotating}
\usepackage{numprint,multirow}
\usepackage{comment}
\usepackage{upgreek}
\usepackage{authblk}

\title{Deep Feature based Cross-slide Registration}

\author[1]{Ruqayya Awan}
\author[1]{Shan E Ahmed Raza}
\author[2]{Johannes Lotz}
\author[2]{Nick Weiss}
\author[1,3,4]{Nasir Rajpoot}
\affil[1]{Department of Computer Science, University of Warwick, UK}
\affil[2]{Fraunhofer Institute for Digital Medicine MEVIS, L\"{u}beck, Germany}
\affil[3]{The Alan Turing Institute, London, UK}
\affil[4]{Department of Pathology, University Hospitals Coventry \& Warwickshire, UK}

\begin{document}
\maketitle

\begin{abstract}
\noindent
Cross-slide image analysis provides additional information by analysing the expression of different biomarkers as compared to a single slide analysis. These biomarker stained slides are analysed side by side, revealing unknown relations between them. During the slide preparation, a tissue section may be placed at an arbitrary orientation as compared to other sections of the same tissue block. The problem is compounded by the fact that tissue contents are likely to change from one section to the next and there may be unique artefacts on some of the slides. This makes registration of each section to a reference section of the same tissue block an important pre-requisite task before any cross-slide analysis. We propose a deep feature based registration (DFBR) method which utilises data-driven features to estimate the rigid transformation. We adopted a multi-stage strategy for improving the quality of registration. We also developed a visualisation tool to view registered pairs of WSIs at different magnifications. With the help of this tool, one can apply a transformation on the fly without the need to generate transformed source WSI in a pyramidal form. We compared the performance of data-driven features with that of hand-crafted features on the COMET dataset. Our approach can align the images with low registration errors. Generally, the success of non-rigid registration is dependent on the quality of rigid registration. To evaluate the efficacy of the DFBR method, the first two steps of the ANHIR winner's framework are replaced with our DFBR to register challenge provided image pairs. The modified framework produces comparable results to that of challenge winning team.

\end{abstract}

\section{Introduction}

Registration often serves as an essential pre-processing step for many medical image analysis tasks. A typical approach for registering two images consists of an optimisation algorithm accompanied by a similarity measure, selected based on the image capture modality. The optimisation algorithm finds the best spatial transformation by maximising the similarity measure which evaluates the correspondence between the images after applying the transformation. In digital pathology, registration has typically been used to capture information from a single modality images IHC stained using different biomarkers and often also including the routine H\&E slides too. Broadly speaking, histology image registration has three main applications: cross slide image analysis \cite{trahearn2017hyper}, multi-modal image fusion \cite{rusu2013multiscale} and 3D reconstruction of serial section \cite{song20133d, xu2015method}. In this work, emphasis has been given to the registration of a stack of consecutive multi-stain histology images for multi-slide image analysis. 

To proceed with the multi-slide analysis, cross-slide alignment of the WSIs for serial sections is a pre-requisite. This is due to the fact that during the slide preparation process, tissue sections cut from the same tissue block will not retain their continuity in the z-axis. Therefore, registering these images is an important step prior to any automated multi-slide co-localisation analysis. The cross-slide registration of histology images is a challenging task due to many reasons: changing structure between the sections, missing parts, tissue folds, broken tissue and even the overall morphology of the tissue could change due to their fragility. There is a need for a registration approach that can perform well under these conditions and is able to align pairs of images in a reasonable time to facilitate the downstream analytical or diagnostic pipeline.

In general, there are two main methods for automatic image registration: intensity-based registration and feature-based registration \cite{bartoli2007image,abdel2017feature}. As the name suggests, intensity-based techniques utilise the raw pixel intensity information in an image without any prior processing or analysis. The correspondence is found by transforming a moving image such that it maximises the similarity measure between the reference image and transformed moving image. Feature-based methods would first identify the key features from the images and then a transformation is estimated by a matching system by utilising the matching features between two images. The choice between the two methods depends on the nature of images. In this work, we are analysing slide images stained with different biomarkers which means that the staining pattern would vary among them. It is unlikely that we will observe any correlation between intensity values of the same tissue region stained with two different biomarkers. Therefore, intensity-based methods generally struggle to compute accurate transformations. For this reason, a feature-based approach is considered more suitable for cross-slide registration.


Cross-section slides mostly have non-linear deformations that cannot be tackled with a global transformation alone. Whereas, non-rigid registration methods are capable to find correspondence by locally transforming a source image. However, registration with Newton-type optimization relies on an initial guess that is close to the optimum to obtain a fast convergence rate and also to prevent converging to local minima. Therefore, a good choice of global transform is required as an initialiser to unify its accuracy with the robustness of the non-rigid transformation. The focus of this study is demonstrating the robustness of data-driven features for estimation of the global transformation aligning multi-stained tissue images. We refer this approach as ``deep feature based registration" (DFBR). To mitigate non-linear deformations, we employed an existing method for non-rigid registration \cite{lotz2019robust, lotz2021high} which has been successfully applied to multi-stain histology images. We present a comparative performance analysis of deep features and the hand-crafted features. Experimentation is done on the COMET dataset and an additional multi-IHC histology dataset which was released by the organisers of a recent challenge contest on non-linear registration.

\section{Literature Review}

CNN features have been shown to outperform handcrafted features by a large margin in several different tasks including registration across the computer vision and medical imaging domains. Several studies have attempted to use deep learning models as feature descriptors in a matching task for medical images. In \cite{wu2013unsupervised, wu2015scalable}, the authors proposed an unsupervised approach for learning most discriminating features which are later integrated into the existing registration tools for prediction of a dense deformation field. This study was tailored to registering MR brain images. In line with this approach, deep features have been used for predicting the rigid transformation for histology images \cite{awan2018deep}. In that work, an autoencoder was trained to generate an output similar to the input to learn a feature representation. Features extracted from the encoder part of a trained autoencoder were used to find the best transformation using gradient descent. 

There is a significant amount of work done with handcrafted features for an alignment task --- \cite{schwier2013registration, wang2014robust, trahearn2014fast, trahearn2017hyper, Nick_Reg_Thesis_2017, solorzano2018whole, borovec2018benchmarking, wodzinski2021multistep} to list a few. Whereas, there were very few CNN based studies on predicting the transformation parameters for a highly deformed pair of images. This is because the known correspondence is needed for training a CNN which is not often available. Also, the trained CNN may not perform comparably well on an unseen dataset. These limitations can be addressed by using an unsupervised approach to some extent and are yet to achieve a significant improvement in terms of registration accuracy. However, there is a wide gap in the extent of utilisation of unsupervised deep learning approach for registration tasks as compared to other image analysis tasks. A major reason is the inability to estimate the transformation while training a network. In 2015, Jaderberg \textit{et al.} \cite{jaderberg2015spatial} proposed a learnable module for applying the spatial transformation to an image, referred to as the `spatial transformer'. Since this module is differentiable, it can be added to any network for end-to-end training. This was not primarily designed for registration purposes; instead, the aim was to enable the CNN to learn features that are invariant to the spatial transformations. After the introduction of the spatial transformer, deep learning gained momentum in designing neural network architectures suitable for registration in an unsupervised learning manner. It has now become a core component of most of the deep learning-based registration methods. Chang \textit{et al.} \cite{shu2018unsupervised} proposed a multi-scale iterative framework for registering microscopic images of serial sections of a mouse brain. A CNN model with a spatial transformer as one of the layers was used to predict the affine transformation. The model was trained to minimise the mean square error between reference and warped moving images. Shengyu \textit{et al.} \cite{zhao2019unsupervised} proposed a deep learning architecture for 3D registration of CT images of the liver and MR images of the brain. The proposed architecture consisted of two sub-networks: one for predicting the affine transformation and the other for predicting the non-linear deformation field. This approach was also applied to the multi-stain histology dataset provided as a part of the ANHIR challenge \cite{borovec2020anhir} and was observed to be the fastest performing method. However, it was not close to the best-performing methods in terms of registration error due to its limited generalisability. It was ranked 6 out of 10 teams who submitted the results. In another study \cite{wodzinski2020learning}, the authors trained a CNN model with good generalisability for predicting the affine transformation in an unsupervised manner. They compared their results with that of SIFT, SURF and  Elastix tool. Their proposed approach didn't outperform other methods in terms of accuracy; however, considering the success rate, the authors claimed that the reported accuracy had been sufficient to perform non-linear alignment successfully. Dwarikanath \cite{mahapatra2020registration} integrated segmentation maps to aid in performing non-linear registration using a self-supervised deep learning-based approach. Segmentation maps were generated by applying $k$-means clustering to concatenated multi-scale feature maps, extracted from a pre-trained segmentation model. The author employed VoxelMorph architecture \cite{balakrishnan2019voxelmorph} for registration and replaced the manual segmentation maps with their fine-grained feature maps. Similar to other non-linear registration methods, this method also required two images to be linearly aligned before its application. 

Most approaches to non-linear registration are designed to work with images that are linearly aligned beforehand. This is because the non-rigid transformation can be more successful when images are linearly aligned. There has been a significant amount of work on non-linear registration. However, the focus has been on monomodal and multi-modal registration of X-ray, CT and MRI images and very few methods have been proposed for histology images. Wodzinski and Muller \cite{wodzinski2020unsupervised} proposed a deep learning based non-rigid registration method, performing comparably to the winning team of the ANHIR challenge contest and is significantly faster than other iterative methods. Their proposed approach employs UNET like architecture, trained in a multi-level unsupervised manner using negative normalised cross-correlation (NCC) as an objective function. However, most data-driven approaches require a large number of training samples to perform highly accurate registration.

Our proposed approach for registering multi-stain images comes under the same umbrella of using CNN as a feature descriptor. Our work is inspired by the work in \cite{yang2018multi} on registering multi-temporal remote sensing images using a CNN, whereby the authors used multi-scale deep features for detection of matching feature points between an image pair. These matching feature points are then used to solve thin-plate spline (TPS) interpolation for image alignment. In our work, we have followed a similar approach for feature description and the correspondence between feature points of two images was found by computing the Euclidean distance measure. Instead of TPS formulation, we used these matching feature keypoints for estimating the affine transformation.

\section{The Proposed Approach}
\begin{figure}
\centering
\includegraphics[scale=0.5]{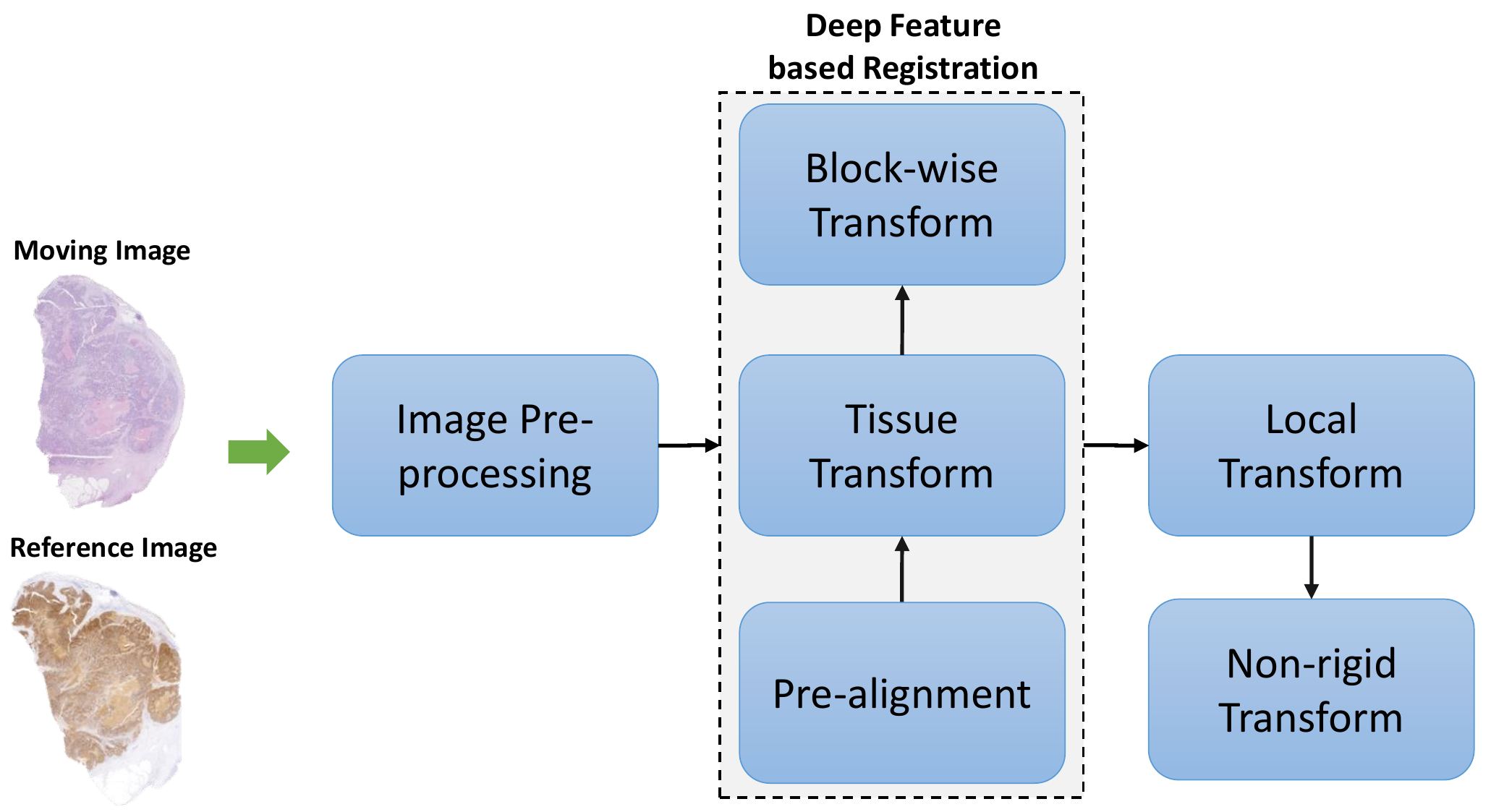}
\caption{Overall pipeline for cross-slide image registration. The pipeline comprises our proposed deep feature based registration (DFBR) method, followed by an existing non-rigid registration method.}
\label{fig:df_reg_pipeline}
\end{figure}

A registration step for any downstream co-localisation analysis workflow should be able to allow a significant spatial overlap between the two images such that the location of corresponding tissue structures can be determined. To this end, we propose a pipeline comprising three main steps: pre-processing, estimation of rigid alignment using our proposed DFBR method, followed by a non-linear registration. During the pre-processing step, we generate a tissue mask for an image pair and modify the input images such that they appear spatially similar. Our DFBR method further contains three sub-modules: \textit{pre-alignment} to perform rough alignment, \textit{tissue transform} estimates the transformation parameters using cropped tissue region and \textit{block-wise transform} refines \textit{tissue transform} by performing feature matching in a block-wise manner. After an image is registered using our DFBR, we observe a slight offset in some cases. To fix this offset, we add a local transform module which is followed by an existing non-linear registration method. The overall proposed pipeline for registration is shown in Figure \ref{fig:df_reg_pipeline}. 

\subsection{Pre-processing Steps}

\subsubsection{Tissue Segmentation}
\begin{figure}
\centering
\includegraphics[scale=0.5]{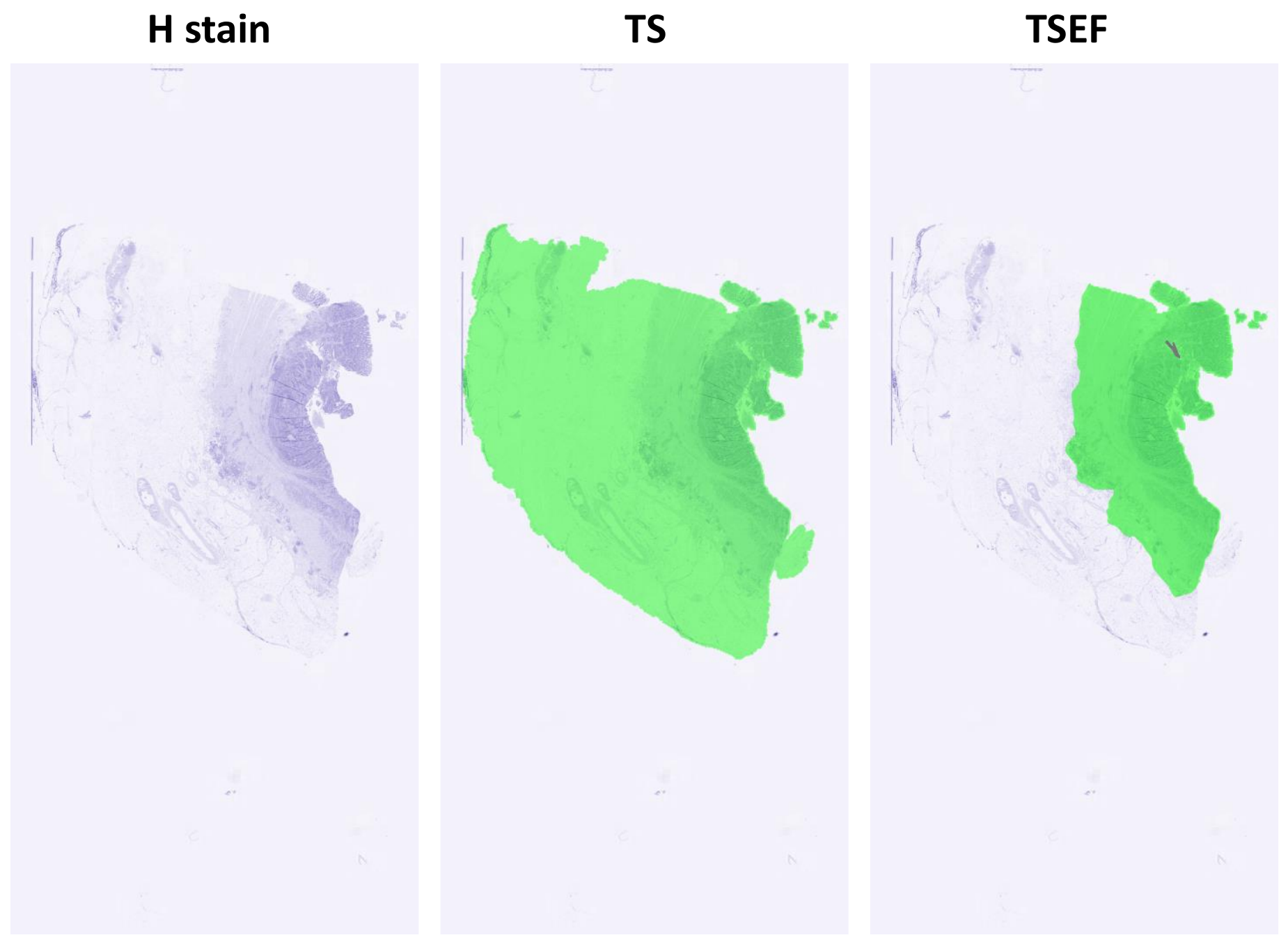}
\caption{An example image with the two types of tissue masks. In TS, all the tissue including fatty region is included whereas in TSEF, fatty region is excluded due to its negative impact on our DFBR registration method.}
\label{fig:tissue_mask_type}
\end{figure}

\begin{figure}
\centering
\includegraphics[scale=0.8]{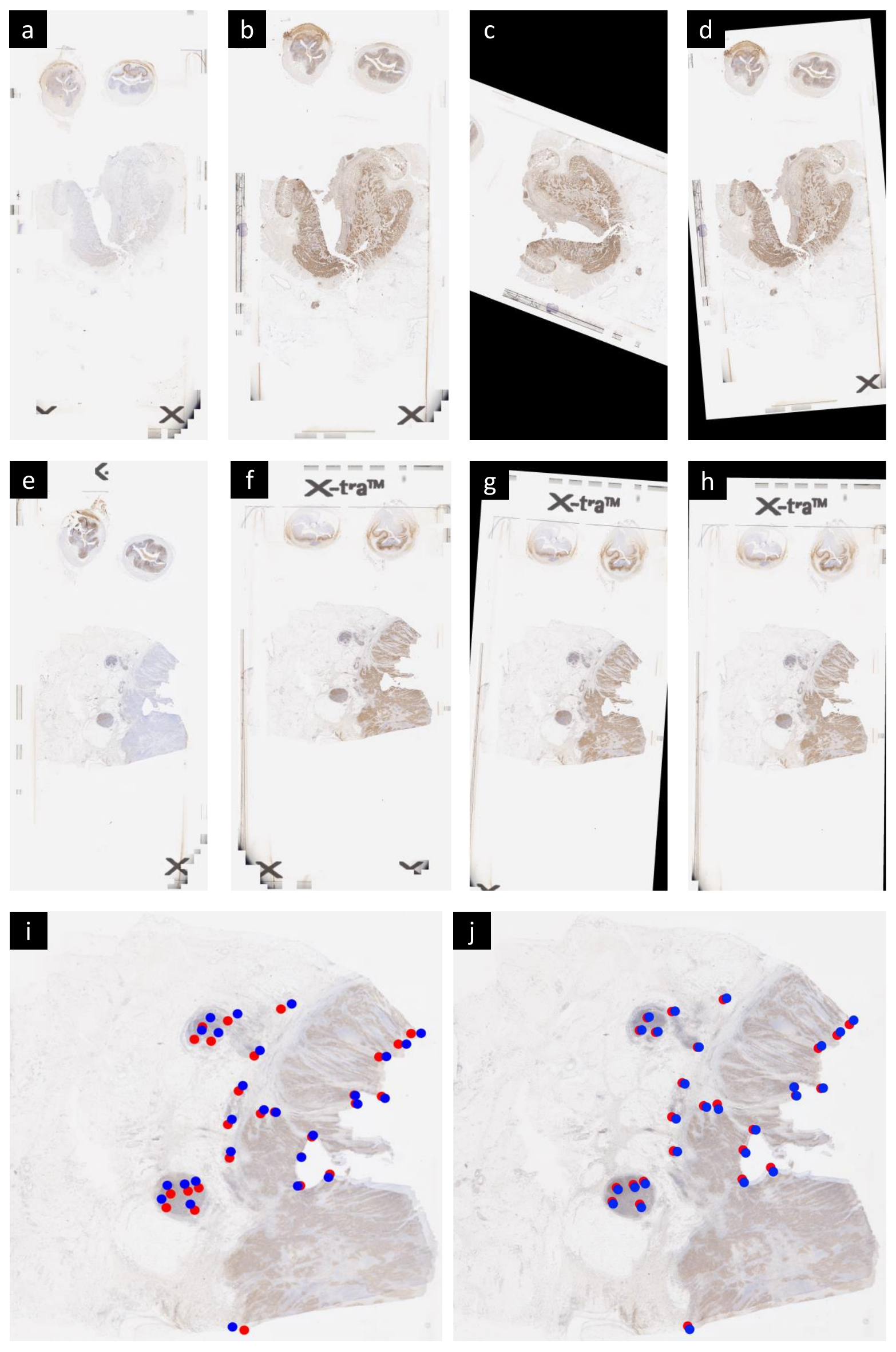}
\caption{A qualitative comparison of the registration accuracy of transformations estimated using segmentation TS and TSEF. First and second columns in the first two rows show reference and moving images, respectively while third and last columns show transformed moving images using segmentation TS and TSEF, respectively. i) shows overlay of e) and g) and j) shows overlay of e) and h). Landmarks are overlaid for the purpose of visualising the offset between the two images.  }
\label{fig:visual_TS_1&2_results}
\end{figure}

We perform tissue segmentation to exclude features from the non-tissue regions. To this end, we train a CNN to generate tissue masks \textbf{TS}, considering all tissue and non-tissue regions as foreground and background, respectively. First, we conduct our DFBR experiments using the TS masks. However, we observe that feature matching points from the fatty region were mostly mismatched. This is because the fatty tissue does not contain adequate texture, thus resulting in low discriminating features. To avoid getting incorrect matching points, we modify our ground truth for tissue segmentation. In the updated masks, we consider the fatty region as a background. In this paper, we refer to this tissue segmentation excluding fat as \textbf{TSEF}. In Figure \ref{fig:tissue_mask_type}, a downsampled version of WSI is shown along with its two different tissue masks to demonstrate the difference between the two. We train a separate CNN for generating these masks so that registration could be carried out using matching points from the active or discriminatory tissue area only (while excluding the ones heavily surrounded by the fatty tissue). Our DFBR method performs better when the transformation matrix was estimated using matching points from the discriminatory tissue area. The quantitative and qualitative comparison of DFBR using TS and TSEF are shown in Table \ref{tab:reg_results_seg_masks} and Figure \ref{fig:visual_TS_1&2_results}. Similar to \cite{Nick_Reg_Thesis_2017}, control tissue is excluded while estimating the transformation parameters.

\begin{table}
\centering
\begin{tabular}{c|c|c|c|}
\cline{2-4}
                                & \textbf{AMrTRE} & \multicolumn{1}{l|}{\textbf{MMrTRE}} & \multicolumn{1}{l|}{\textbf{AMaxrTRE}} \\ \hline
\multicolumn{1}{|l|}{TS} &      0.0090    &   0.0044   &    0.0208        \\ \hline
\multicolumn{1}{|l|}{TSEF} &      \textbf{0.0063}    &   \textbf{0.0041}   &    \textbf{0.0176}    \\ \hline
\end{tabular}
\caption{Comparative results of the proposed DFBR method using two different tissue segmentation masks. The exclusion of fatty areas significantly minimised the registration error. These results are generated using the COMET dataset.
}
\label{tab:reg_results_seg_masks}
\end{table}

\subsubsection{Input Image Pre-processing}
CNN models trained for a classification problem are known to be biased towards texture rather than the colour of the input image \cite{geirhos2018imagenet}. Based on this observation, we can perform feature matching using original images without any pre-processing. To explore it with our registration pipeline, we experiment with four different versions of the input images: original RGB images, greyscale images, blue ratio and Haematoxylin (H) stain images. To unify the appearance of two images, we perform histogram matching as a normalisation step. In histogram matching, the histogram of an image is modified to be similar to that of another image. It is performed for all pairs of images except those with RGB values. In our experiments, an image with high entropy is considered as a reference image and the histogram of an image with low entropy is matched to the reference image. Since most pre-trained CNN models accept input images with 3 channels, greyscale, H and blue ratio images were stacked as the colour channel. In our experiments, we find greyscale images to perform better as compared to other input versions, as shown in Table \ref{tab:reg_results_input}.

\begin{table}
\centering
\begin{tabular}{c|c|c|c|}
\cline{2-4}
                                & \textbf{AMrTRE} & \multicolumn{1}{c|}{\textbf{MMrTRE}} & \multicolumn{1}{c|}{\textbf{AMaxrTRE}} \\ \hline
\multicolumn{1}{|c|}{Blue Ratio}    &     0.0145    &   0.0058   &    0.0285  \\ \hline
\multicolumn{1}{|c|}{RGB}           &     0.0077    &   0.0043   &    0.0190  \\ \hline
\multicolumn{1}{|c|}{H stain}       &     0.0070    &   0.0044   &    0.0185  \\ \hline
\multicolumn{1}{|c|}{Greyscale}     &     \textbf{0.0063}    &   \textbf{0.0041}   &   \textbf{0.0176}  \\ \hline
\end{tabular}
\caption{Demonstration of the efficacy of our DFBR approach using different versions of input image pairs. Greyscale images are shown to outperform other pre-processed input images. These results are generated using the COMET dataset.}
\label{tab:reg_results_input}
\end{table}

\subsection{Alignment}
Broadly speaking, the tissue alignment is performed in three main steps: pre-alignment, rigid alignment and non-rigid alignment. The output registered image generated in each step is given as an input to the next step along with reference image. All these steps are discussed in detail in the following sections. 

\subsubsection{Pre-alignment}
In this step, rough estimates of translational and angular offsets are computed. Since CNN features are not rotation invariant, this step is key to performing deep feature matching successfully. First, we estimate the translation offset by finding a centre of mass (COM) for an image pair. The COM is a vector of \textit{x} and \textit{y} coordinates and is computed from the inverted greyscale intensity values of the tissue region only. The difference between the COM values of a pair of images is used to obtain a translation matrix. This matrix transforms the moving image such that its COM is at the same position in the coordinate system as that of a reference image. Next, we find a rotation matrix by an exhaustive search strategy resulting in a maximum overlap between the tissue masks of an image pair. Once a moving image is roughly aligned to a reference image, we crop the tissue regions from an image pair using their tissue masks. We determine a bounding box that includes the tissue region of both images. In the following steps, we only use tissue regions instead of the whole image for registration.   

\subsubsection{Deep Feature based Registration}
\begin{figure}
\centering
\includegraphics[scale=0.6, angle=90]{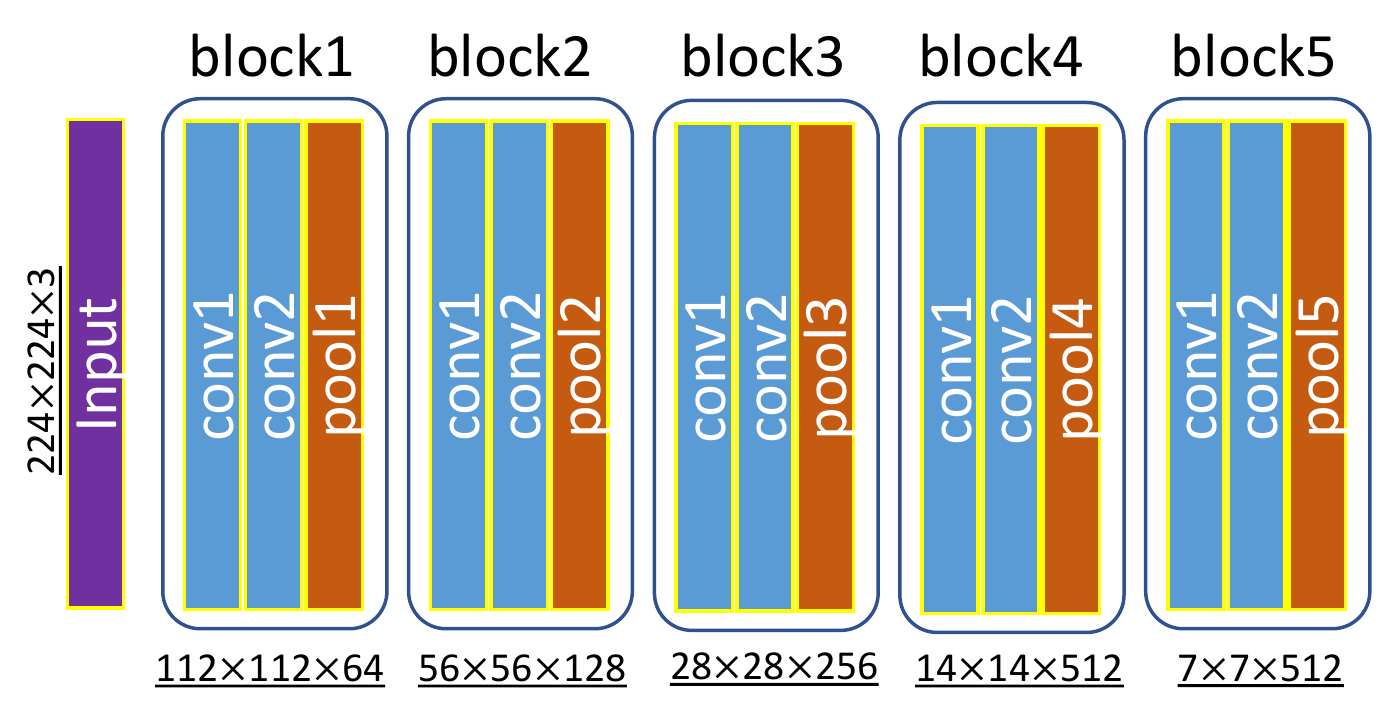}
\caption{A partial architecture of the VGG16 model, used for feature extraction in our DFBR module.}
\label{fig:vgg_architecture}
\end{figure}

The objective of our feature based alignment step is to refine the alignment between reference and pre-registered moving images by registering their feature points. We present data-driven features extracted using a pre-trained VGG-16 model. The partial architecture of VGG-16 that we use for feature extraction is shown in Figure \ref{fig:vgg_architecture}. We extract multi-scale features for an image pair and find the matching pairs by considering the feature points with a small feature distance. 

\begin{figure}
\includegraphics[width=\linewidth]{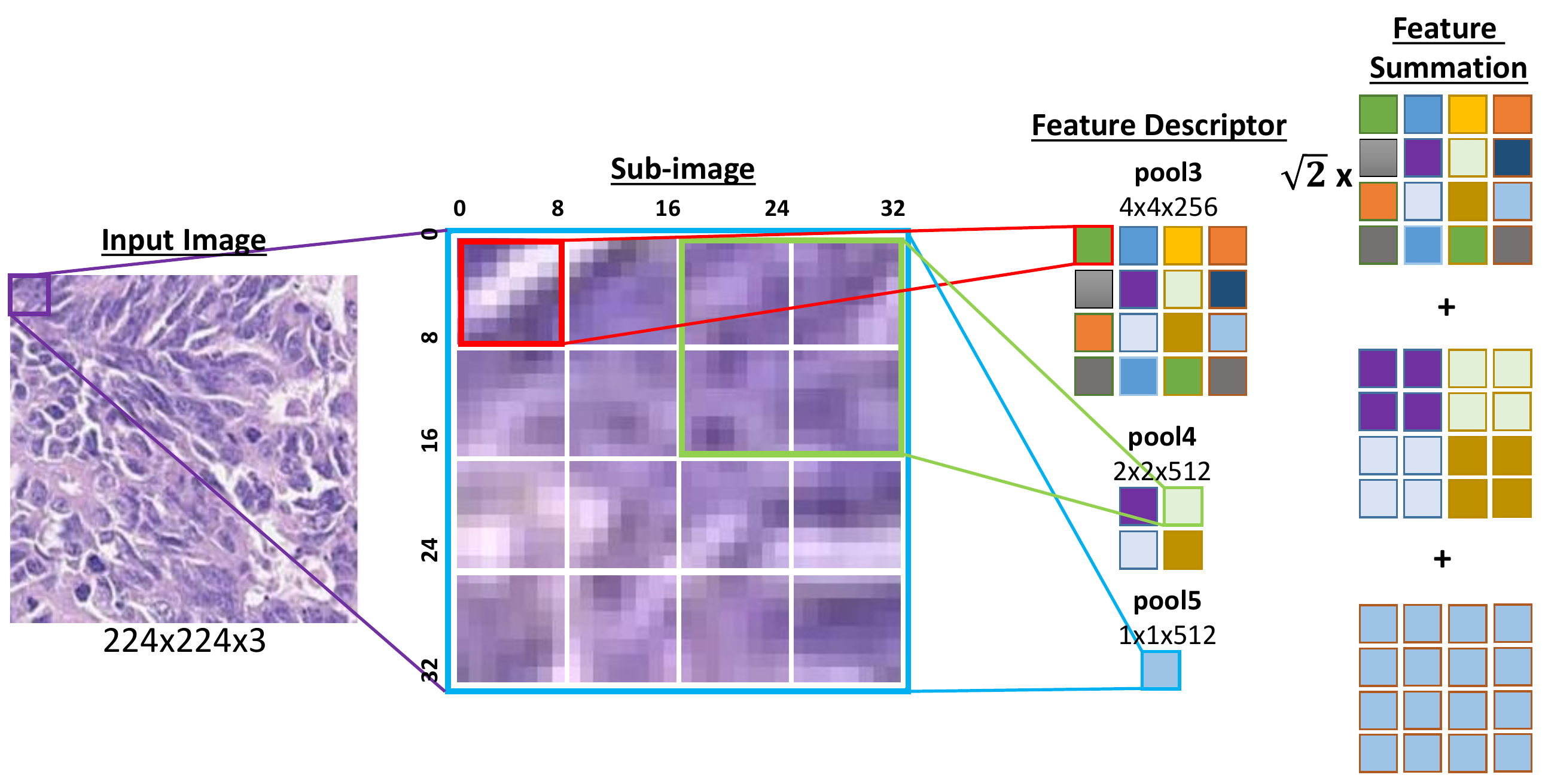}%
\caption{Demonstration of feature summation for a sub-image.}
\label{fig:feature_summation}
\end{figure}

\textbf{A. Feature Descriptors:} In a handcrafted feature based method for registration, the first step is to identify/detect the feature points containing distinctive information such as a corner, blobs, edges etc. It is then followed by a feature extraction step which involves the computation of descriptors. Descriptors describe the properties of regions centred at the distinctive feature points. These feature descriptors are used to find correspondence between two images in a feature matching step. In our proposed feature based method for alignment, the feature detection step doesn't exist. Instead, an image is divided into a grid and a feature descriptor is computed for every grid cell in a sliding window fashion. Similar to \cite{yang2018multi}, our feature descriptor is formed by deep features extracted from three different layers of a VGG-16 model, pre-trained for ImageNet classification. Since the bottom fully connected layers are removed, the partial model can take input images of any size (spatial dimensions multiples of 32), with larger images increasing the computation time. Our experiments are conducted with an input size of $224\times224$ pixels. Features extracted from the bottom three pooling layers (pool3, pool4 and pool5) are used to build descriptors. Each of these layers corresponds to receptive fields of different sizes as shown in Figure \ref{fig:feature_summation}. pool3 is considered as a reference feature extractor layer; therefore, the features of pool4 and pool5 are mapped to those of pool3. Following are some annotations that we use in this section: $F_{i}^{j}$ refers to the features of an image $j$ extracted from a pooling layer $i$. For example, pool3 layer features of a reference image $R$ is denoted by $F_{3}^{R}$. 


\begin{itemize}
  \item \textbf{pool3} outputs features $F_{3}$ of dimension $28\times28\times256$. Each feature has a receptive field of size $8\times8$, dividing input image into a $28\times28$ grid. The center of each grid cell is considered a feature points for the respective descriptor.  
  \item \textbf{pool4} outputs features $F_{4}$ of dimension $14\times14\times512$. Each feature has a receptive field of size $16\times16$ and therefore it is shared by four feature points. 
  \item \textbf{pool5} outputs features $F_{5}$ of dimension $7\times7\times512$, each feature has a receptive field of size $32\times32$ and therefore it is shared by 16 feature points. 
\end{itemize}

\noindent
Each of these descriptors is normalised to unit variance. 

\textbf{B. Feature Matching:} 
Once feature descriptors are formed for an image pair, we compute the Euclidean distance between their feature points. The distance metric for features of layer \textit{i} is computed as

\begin{equation*}
D_{i}(p^{R}, p^{M}) = distance_{euc}(F_{i}^{R}, F_{i}^{M})
\end{equation*}

\noindent
where $i \in{[3,4,5]}$. Each value in a feature distance matrix for pool3 relates to an individual feature point, which is not the case with distance matrices for pool4 and pool5 features. Each distance value in $D_{4}$ and $D_{5}$ corresponds to 4 and 16 feature points, respectively. Therefore, we replicate each distance value in $D_{4}$ 4 times. Similarly each distance value in $D_{5}$ is replicated 16 times. After replication, $D_{3}$, $D_{4}$ and $D_{5}$ are added, with some weight given to $D_{3}$ due to a smaller number of features. Feature distance between two feature points $p^{R}$ and $p^{M}$ is computed as

\begin{equation*}
D(p^{R}, p^{M}) = \sqrt{2}D_{3}(p^{R}, p^{M}) + \emph{replicate}(D_{4}(p^{R}, p^{M}), 4) + \\ \emph{replicate}(D_{5}(p^{R}, p^{M}), 16)
\end{equation*}

The quality of each matching feature point is determined by the difference between the smallest and second smallest Euclidean distances. The greater the difference, the better the quality.



\noindent
There are two conditions for matching point $p^{R}$ to point $p^{M}$
\begin{enumerate}
  \item $D(p^{R}, p^{M}) < D(:, p^{M})$; which means there shouldn't be any other feature distance smaller than $D(p^{R}, p^{M})$.
  \item The quality of matched features should be greater than a threshold value which is computed automatically for each image pair. A threshold value is set such that \textit{U} number of pairs of matching points are selected. 
\end{enumerate}

Using the above steps, we find matched feature points between any two images. Similar to \cite{yang2018multi}, we select \textit{U} = 128 pairs of matched points. We then use the matched feature points as control points for estimating the transformation parameters. Using a least squared approach for finding a rigid transformation that best align these matched points. To use this method for histology image registration, we apply this step twice. First, we apply it to the whole tissue region for finding the best matched points globally and is referred as the `tissue transform'. Secondly, we divide the tissue regions into four parts and apply feature matching method to each part individually. This is referred to as the `block-wise transform'. It results in higher number of matching feature points and is likely to further improve the alignment. Also, block-wise feature matching step can be applied in a parallel fashion to speed up the process.


\subsubsection{Local Alignment}
\begin{figure}
\centering
\includegraphics[scale=0.5]{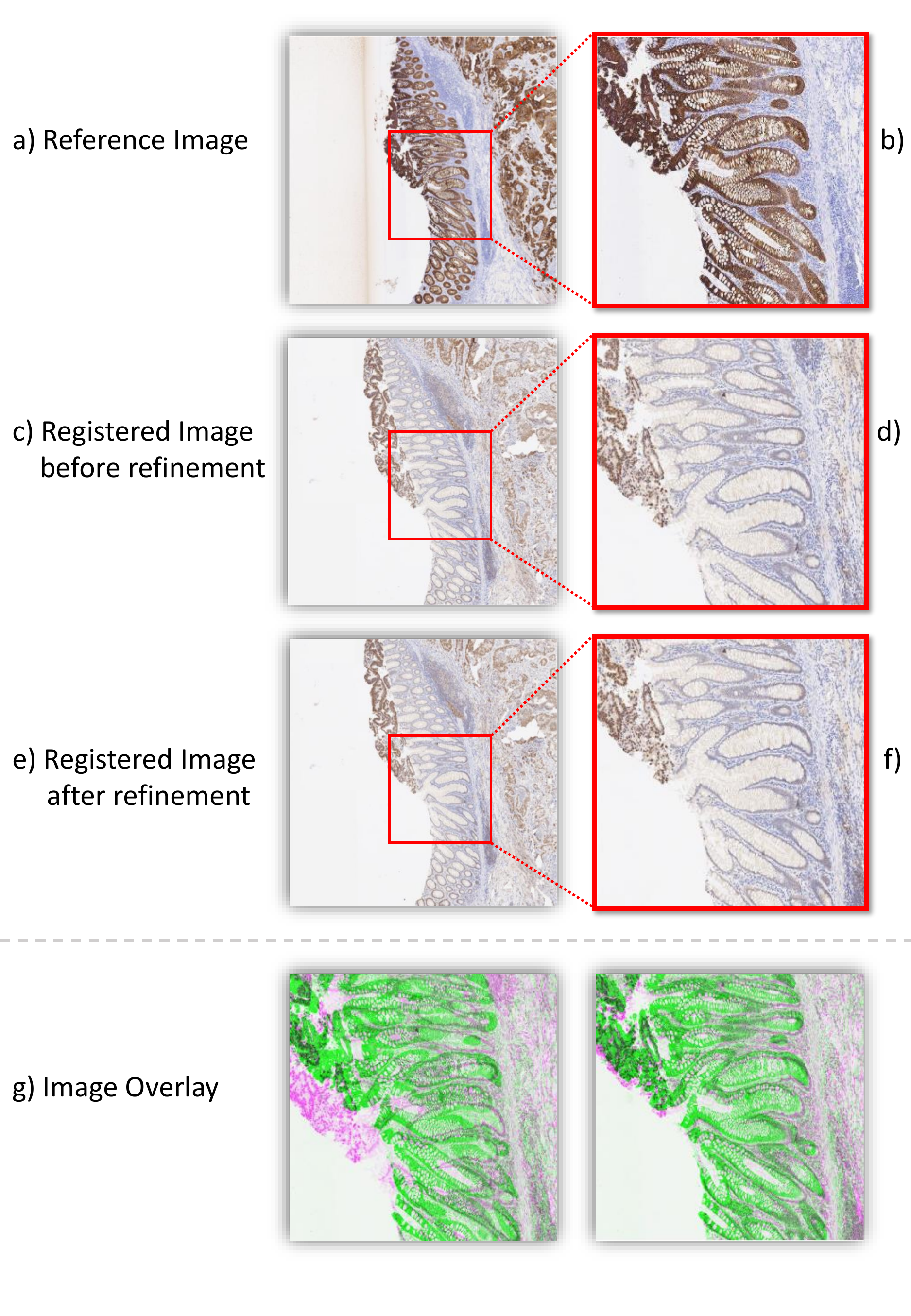}
\caption{Example images of the local phase-based refinement method integrated into our visualisation tool (see section \ref{visual_tool}) to fix translation offset. In bottom row g) we show overlaid false colour images before (left) and after refinement (right). Reference and registered moving images are shown in green and purple colours, respectively.}
\label{fig:phase_refined}
\end{figure}

On visualising a registered image alongside its corresponding reference image at both low and high resolution, we observe global and local translation offsets, respectively. The global offset is likely to exist due to the offset in patches containing the matching key-points (center of the patch). On viewing an image pair at a higher resolution with a minor global offset, a local translation offset between reference and registered images can be observed. This is likely to persist due to the non-linear deformations in some parts of the images. Figure \ref{fig:phase_refined} shows an image overlay of reference and registered moving tissue sections along with their corresponding landmark points. The non-overlapping landmarks demonstrate the local offset between the two images. With landmark points, it can be observed that some parts of the image pair are properly aligned where landmarks of the two images are overlapping, while regions with distant corresponding landmarks indicate a local translative movement between the two images. To fix a global offset, we employ the phase correlation method after applying DFBR to determine a shift between two images at a magnification 0.3125$\times$. This method can be applied to the greyscale or H stain images. However, we found it performing better when applied to tissue masks. We also integrate this method in our visualisation tool (section \ref{visual_tool}) to fix the local shift. During visualisation, the user can fix the offset by clicking on the `Fix Offset' button. An example of an image pair before and after local refinement is shown in Figure \ref{fig:phase_refined}.

\subsubsection{Non-Rigid Alignment}
\begin{figure}
\centering
\includegraphics[scale=0.5]{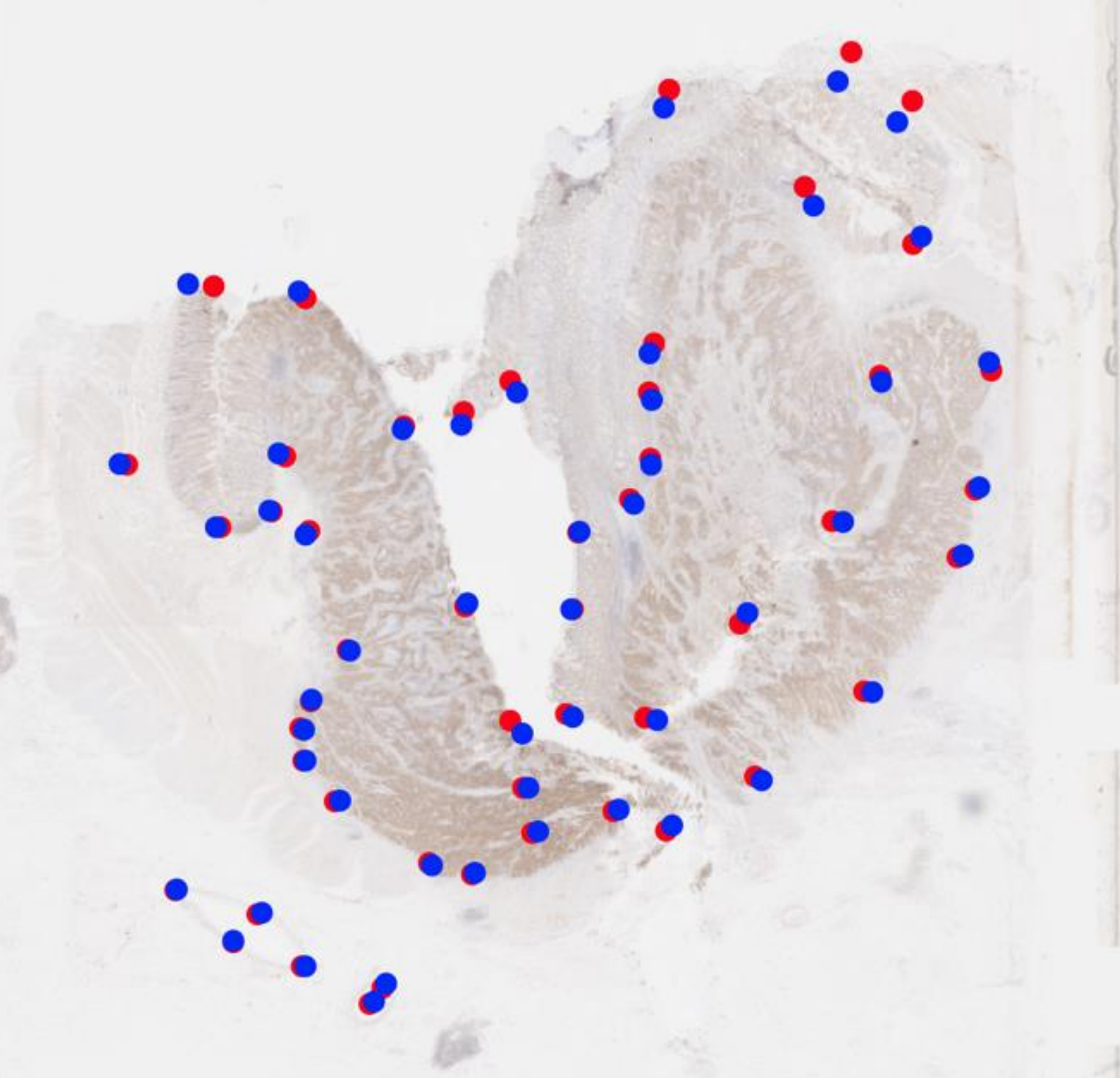}
\caption{Cropped tissue region of a registered image overlaid on the corresponding region of a reference image with their ground truth landmarks. By looking at the landmarks, it is quite evident that some regions are not perfectly registered. This is due to the non-linear deformations in the tissue. }
\label{fig:non-linear_deformation}
\end{figure}

Once an image pair is registered using rigid transformation, it is often the case that some of the tissue areas are not accurately aligned to that of a reference image. As an example, an overlay image of reference and registered images is shown in Figure \ref{fig:non-linear_deformation}. The landmarks are also displayed over images. It can be observed that most of the landmarks are accurately overlapping while there are some landmarks that are not registered well. This is due to the fact that the tissue slices are so thin and fragile that the slide preparation step is likely to introduce non-linear deformations and artefacts such as tissue folds, tissue stretching and compression and even torn/missing tissue parts \cite{pichat2018survey}. The presence of non-linear deformations makes the registration process more challenging. Since these artefacts change the morphology of the tissue, none of the rigid registration methods can tackle such deformities. Therefore, a non-rigid registration approach is applicable in such scenarios. It should be noted that the non-rigid transformation can be applied successfully when two images are linearly aligned. Otherwise, it can introduce further artefacts in the transformed image.  

There are many existing non-rigid registration methods in the literature that have been used for histology image alignment. In the ANHIR challenge, a number of non-rigid methods were evaluated using a benchmark dataset. We apply a non-rigid algorithm \cite{lotz2019robust} proposed by the winner of the ANHIR challenge (MEVIS group) on top of our DFBR method to further improve the registration accuracy. Their proposed method as used in the challenge comprises three steps: pre-alignment, parametric registration and non-parametric registration. In the parametric registration step, affine transformation is estimated based on the greyscale intensity values using the NGF distance measure \cite{haber2006intensity} and a Gauss-Newton optimization. The estimated transformation parameters are then set as an initialiser for the non-parametric dense registration. The displacement field is estimated by minimising their proposed objective function comprising a normalised gradient field (NGF) distance measure along with a curvature regulariser. L-BFGS is used for optimisation. The efficacy of non-parametric registration is highly dependent on the accuracy of the initialiser transform. We experimented with their method after replacing their first two steps with our DFBR method.



\section{Visualisation Tool}
\label{visual_tool}
\begin{figure}
	\centering
	\includegraphics[scale = 0.3]{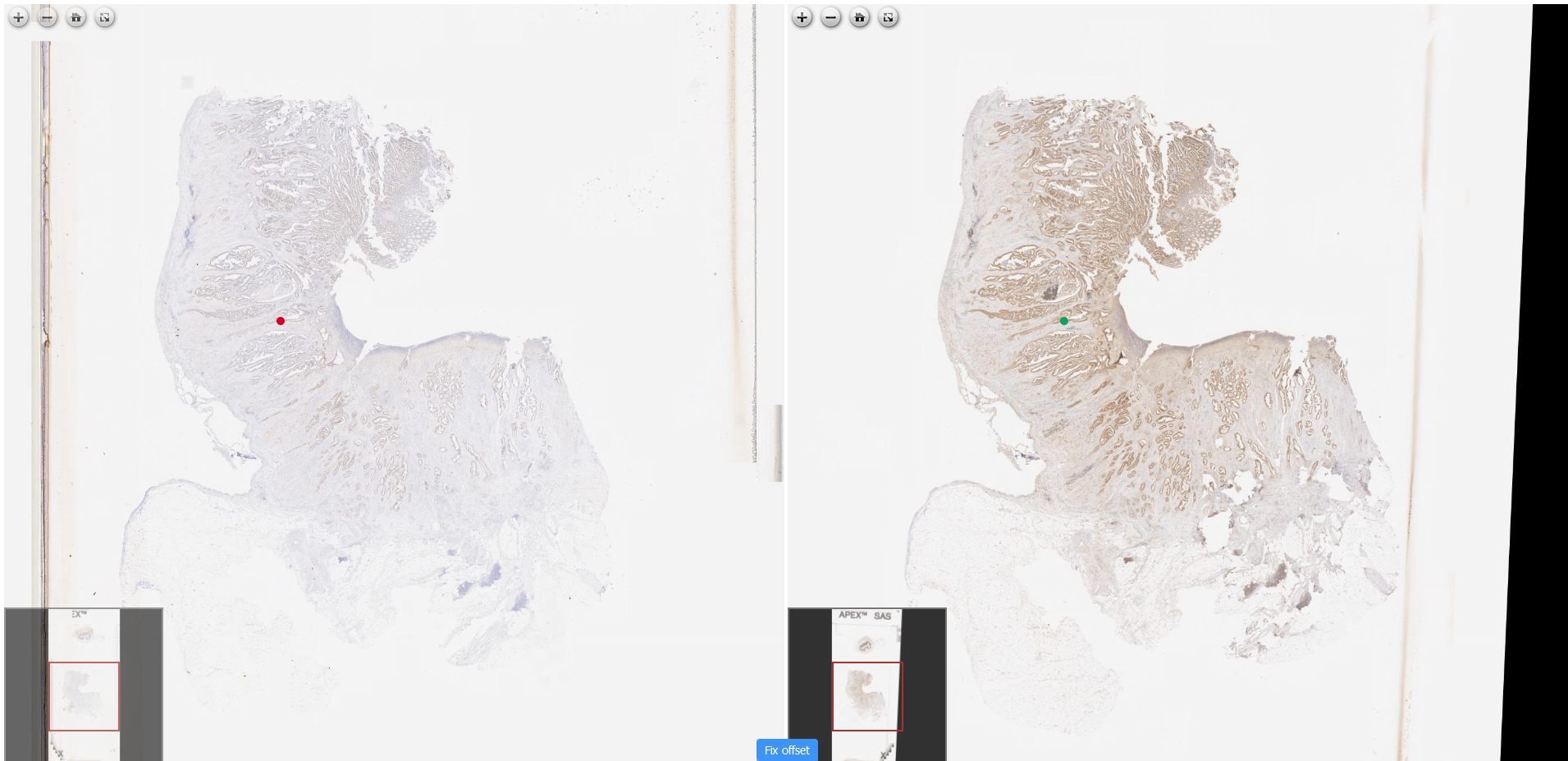}%
	\caption{Visualisation Tool (Interface-1). Reference and registered moving images are displayed on the left and right side panels, respectively. Zooming and panning on these panels are interlinked. It allows the user to visualise corresponding regions of two registered WSIs at any resolution level. }
	\label{fig:visual_tool}
\end{figure}

To visualise registered images while being able to zoom in and out and pan across a WSI, we develop two different web-based tools with a split screen, left panel for displaying the reference image and the right panel for the registered moving image. On each split screen, a dot pointer with a different colour is shown which changes its position with the mouse movement. The regions pointed by these points on two screens are significantly helpful in visually estimating the performance of registration method. A screenshot of the interface is shown in Figure \ref{fig:visual_tool}. 


\textbf{Interface-1} is developed to display results of our deep feature based method. The input to this tool comprises three directory paths: to reference and moving images and the pre-computed transformation parameters. Registration is applied to the tiles on the fly as they are viewed. Therefore, there is no need to generate transformed WSI in a pyramidal format. Since registration is performed at the lowest resolution, we observe a translation offset between image pairs. It may also be due to the non-linear transformation. In this case, a partial tissue is distorted such that no global rigid transformation can fix the whole tissue. We deal with such misalignment by using phase correlation method. We add a button to the interface for the user to fix the offset. Once the offset is computed, it is applied to every FOV as user zoom or pan through the slide. 

\textbf{Interface-2} is developed to display results of non-rigid registration method. It is purely a visualisation tool with no operation on tiles at the back-end while user is visualising them. This is developed to visualise reference and registered images saved in a pyramidal format. After non-rigid registration, we save the registered images as multi-scale images and use this interface to assess the performance visually. This interface takes two inputs only, path to reference and transformed moving images. 


\section{Datasets \& Performance Measures}
We evaluate the performance of our proposed DFBR method using two multi-stain datasets: the COMET and ANHIR datasets. A detailed description of the datasets and evaluation metrics is presented in the following subsections.

\subsection{Datasets}
\begin{figure}
\centering
\includegraphics[scale=0.6]{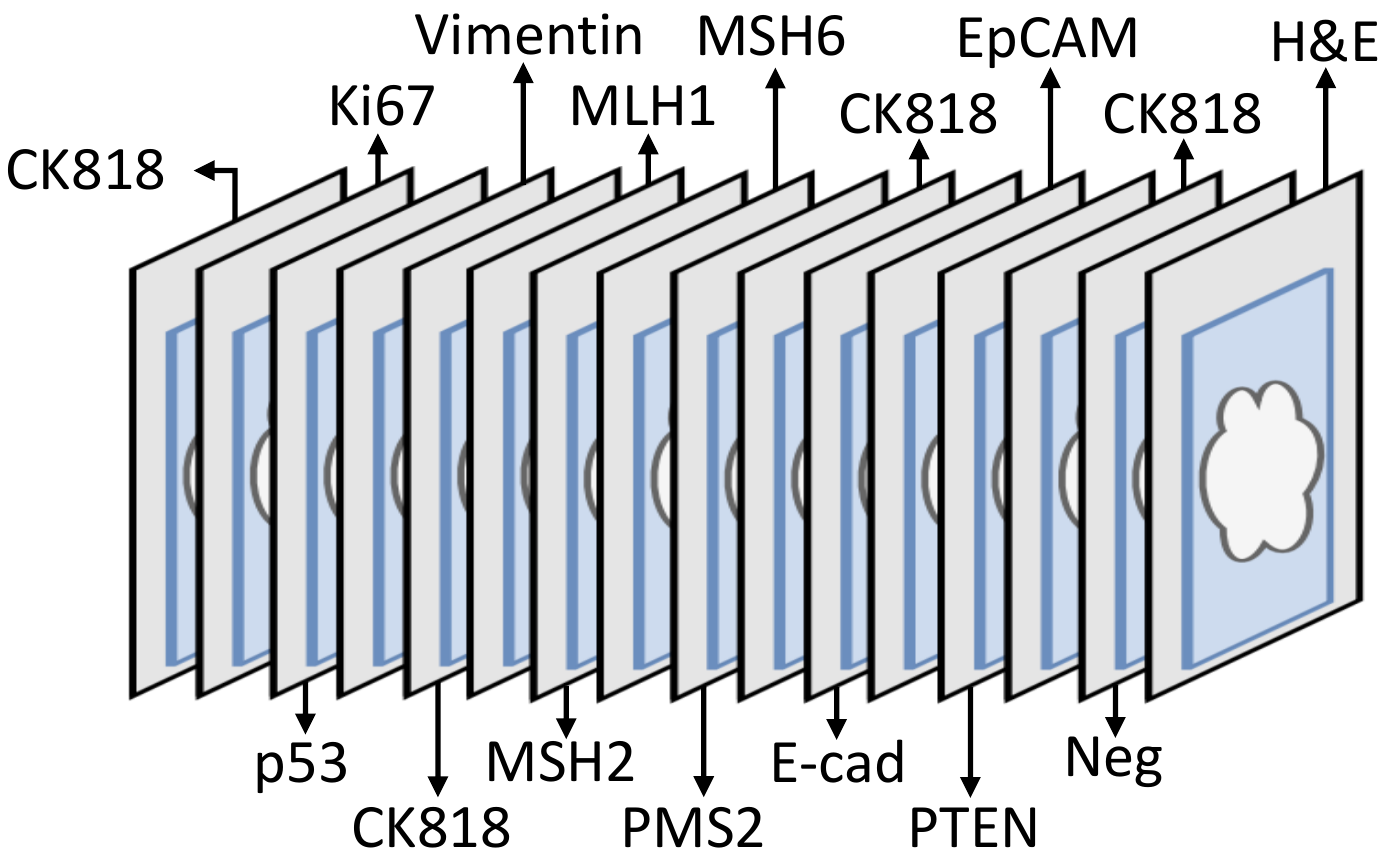}
\caption{An example to illustrate the exact sequence of sections cut from a tissue block in the COMET dataset.}
\label{fig:section_sequence}
\end{figure}

\begin{figure}
\centering
\includegraphics[width = \linewidth]{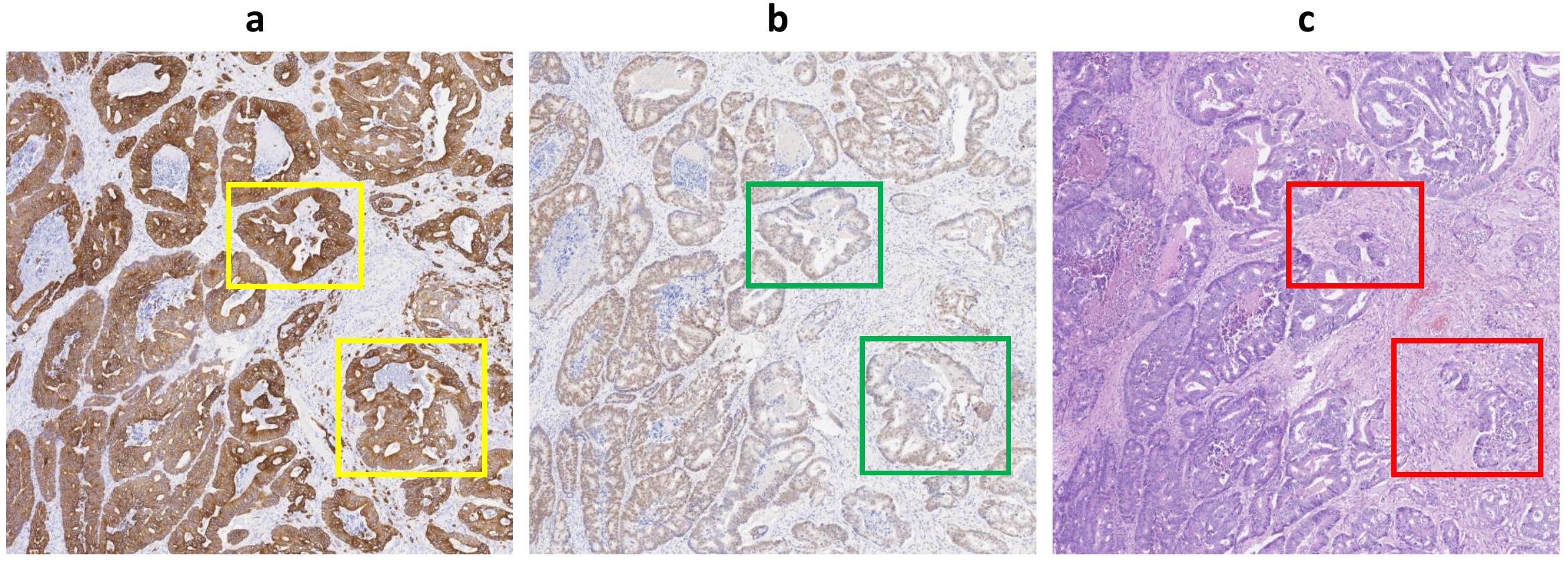}
\caption{Similarities and dissimilarities in the tissue architecture among spatially corresponding visual fields extracted from a) CK818, b) MLH1 and c) H\&E stained images. a and b are taken from the consecutive slides (5$\upmu$m apart), hence has more similarities as compared to a and c or b and c (more than 50$\upmu$m apart).    }
\label{fig:VF_similarity_differences}
\end{figure}

COMET dataset is obtained from the University Hospitals Coventry and Warwickshire (UHCW) NHS Trust in Coventry, UK. This dataset comprises WSIs of 86 cases, taken from different patients. There are 16 slides per case, each scanned using the Omnyx VL120 scanner at 0.275 microns/pixel. These slides are stained with different stains: CK818, Ki67, p53, Vimentin, MMR (MLH1, MSH2, MSH6 and PMS2), E\-cadherin, EpCAM, PTEN and H\&E and their exact sequence of staining is shown in Figure \ref{fig:section_sequence}. We select a set of 7 cases and present the quantitative evaluation of registration methods using them. We consider six slides per case, involving MMR prediction from H\&E and CK818 images for the end purpose of MSI prediction. Therefore, we perform registration of MMR slides with CK818 and H\&E slides. The alignment of H\&E wrt CK818 is challenging as there were 10 sections in between them, including the MMR markers' sections. While CK818 and MMR biomarker slides are the consecutive sections sliced in a given order: CK818, MLH1, MSH2, MSH6 and PMS2. These IHC stained slides tend to be highly correlated in terms of tissue structures which is not the case with H\&E since it is around 50$\upmu$m away from the given IHC slides. Therefore, a significant variability in the tissue structures exists among the H\&E and IHC stained slides, as shown in Figure \ref{fig:VF_similarity_differences}. For evaluation, we consider 15 pairs of sections per case for registration: aligning MMR markers w.r.t CK818 (4 pairs), aligning all IHC slides w.r.t H\&E (5 pairs) and all possible combinations of MMR biomarker slides (6 pairs). This results in 105 pairs in total against 7 cases. To evaluate the performance of registration methods, we manually defined landmarks on all the images selected for evaluating registration performance. ImageJ tool is used for annotating significant tissue structures with landmarks.

\begin{figure}
\centering
\includegraphics[scale=0.6]{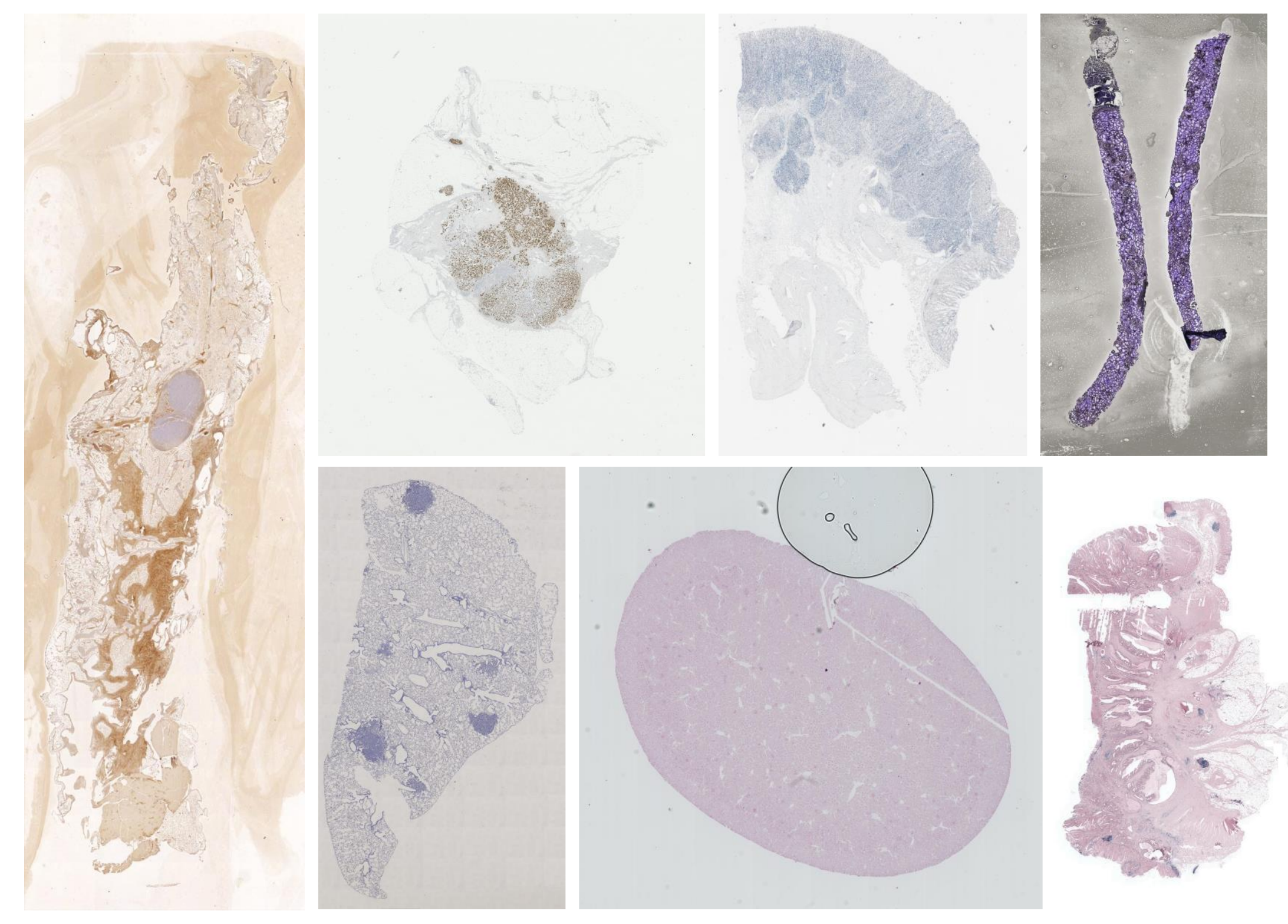}
\caption{Example images from the ANHIR dataset. Each image is stained with a different stain.  }
\label{fig:anhir_dataset}
\end{figure}

The ANHIR dataset is a public dataset, made available by organisers of the ANHIR challenge. It comprises 8 different tissue types stained with 18 different stains, hence making it a challenging dataset. Example images are shown in Figure \ref{fig:anhir_dataset}. There are 230 training and 251 testing pairs for registration. For more details on this dataset, readers are referred to the challenge paper \cite{borovec2020anhir}. The landmarks are provided for the training set while landmarks of reference images in test image pairs are kept private by the organisers. The evaluation of a registration method can be performed by uploading the results on the challenge portal.


\subsection{Performance Measures} 
To measure the quality of registration, we compute the target registration error (TRE) for image pairs in the test set $\mathcal{T}$. For an image pair $j$, we compute the distance-based error measure using the following formula

\begin{equation*} \label{eq:1}
TRE(R_{j}, M_{j}^{'}) = ||R_{land_{j}}, M_{land_{j}}^{'}||_{2}
\end{equation*}

\noindent where $R$ and $M^{'}$ belong to an image pair $j$ and represent reference and transformed moving images, respectively. $R_{land}$ and $M_{land}^{'}$ denote landmarks of reference and transformed moving images, respectively. The registration error is normalised by the length of the reference image diagonal. 

\begin{equation*} \label{eq:2}
rTRE(R_{j}, M_{j}^{'}) = TRE(R_{j}, M_{j}^{'})/hypot(R)
\end{equation*}

\noindent
where $hypot(R) = \sqrt{w^{2} + h^{2}}$ and $w$ and $h$ denote width and height of the reference image, repectively. Above equation for computing rTRE generates a list of values for an image pair $j$ which we aggregated by taking their median. Overall registration error for $\mathcal{T}$ is computed by either taking mean or median of the aggregated rTRE, namely as the median of median rTRE (MMrTRE) and the average of median rTRE (AMrTRE). We also report the average of the maximum rTRE (AMaxrTRE). 

We also test the robustness of registration results by comparing the transformed landmarks with the initial landmarks before any alignment. We compute the robustness in a similar way defined by the ANHIR challenge organisers which is the relative number of successfully transformed landmarks. Any given landmark pair $R_{land^i}$ and $M_{land^i}^{'}$ is considered to have been registered successfully only if the distance between them is smaller than the difference between $R_{land^i}$ and landmark of moving image before registration $M_{land^i}$. Robustness for an image is computed by counting the number of successfully registered landmarks divided by the total number of landmarks for that image. Robustness over the whole dataset is computed by taking the mean over all the image pairs' robustness.

\section{Experimental Results}

\subsection{Experimental Setup}
We consider different scales of information in our registration pipeline. For the COMET dataset, downsampled WSIs at magnification 0.15625$\times$, 0.3125$\times$ and 0.625$\times$ are used for pre-alignment, DFBR including local transform and non-rigid alignment, respectively. For the ANHIR dataset, downsampled WSIs are provided rather than WSIs with varying downsampling rates for different datasets. We rescale the images to 5\% and perform our DFBR registration, followed by a non-rigid registration. For non-rigid registration, we use the same parametric values as the challenge winner. These parametric values are reported in Table \ref{tab:non-rigid_params}. The python based library Keras was used for the extraction of CNN features while the rest of the implementation is in both Python and MATLAB. The implementation of our visualisation tool is carried out in Python and JavaScript. We used OpenSeadragon, an open-source viewer for this tool.

\begin{table}
\centering
\begin{tabular}{|c|c|}
\hline
\textbf{Parameters}     & \textbf{Values} \\ \hline
Number of levels        & 7*               \\ \hline
Maximum image dimension & 8000 pixels     \\ \hline
NGF $\varepsilon$                    & 1.0             \\ \hline
Regulariser parameter $\alpha$  & 0.1             \\ \hline
Control point grid size & 257 $\times$ 257 \\ \hline      
\end{tabular}
\caption{Parameters of non-rigid registration method: $\alpha$ is a regularisation parameter to control the smoothness in deformation while $\varepsilon$ controls the sensitivity of NGF to the noise, and * indicates the parameter that differs in the COMET dataset. We performed registration at a total of 6 levels for the COMET dataset.}
\label{tab:non-rigid_params}
\end{table}


\subsection{Result Summary}
\noindent
\textbf{COMET:} As discussed previously, we observe mismatched feature points using our DFBR method, mainly in the fatty tissue area. Therefore, we replace our tissue mask TS with TSEF. The removal of mismatched feature points from the fatty area improves our results, as shown in Table \ref{tab:reg_results_seg_masks}. Table \ref{tab:reg_results_input} shows results obtained with different versions of input images: original RGB images, greyscale images and H stain images. As our initial experiments show greyscale images to perform better, greyscale images are used in all our experiments.

\begin{figure}
\centering
\includegraphics[width=\linewidth]{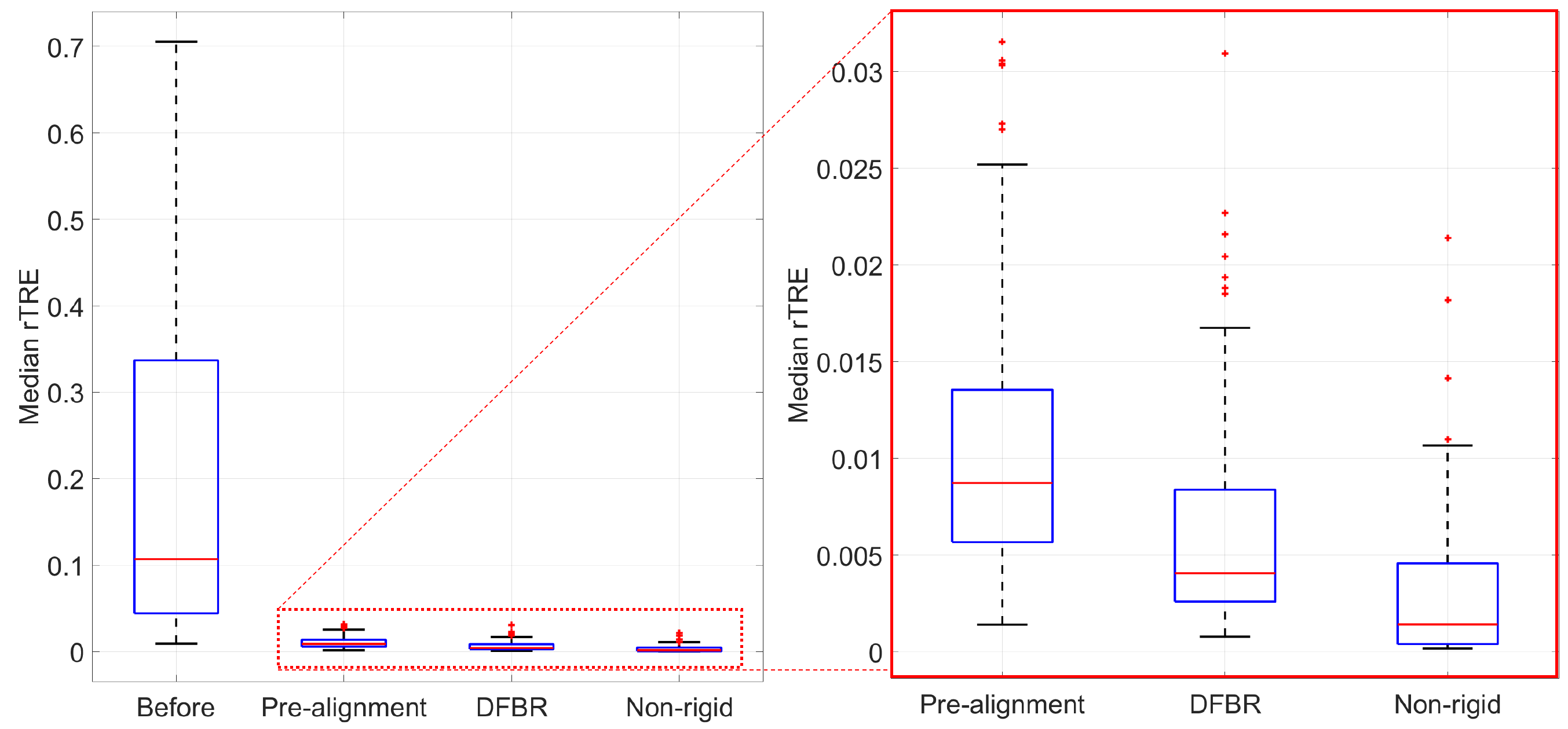}
\caption{Demonstration of the median rTRE before and after applying each registration module using a box plot. The median rTRE is computed for the COMET dataset. }
\label{fig:box_plot_comet}
\end{figure}

\begin{figure}
\centering
\includegraphics[width=\linewidth]{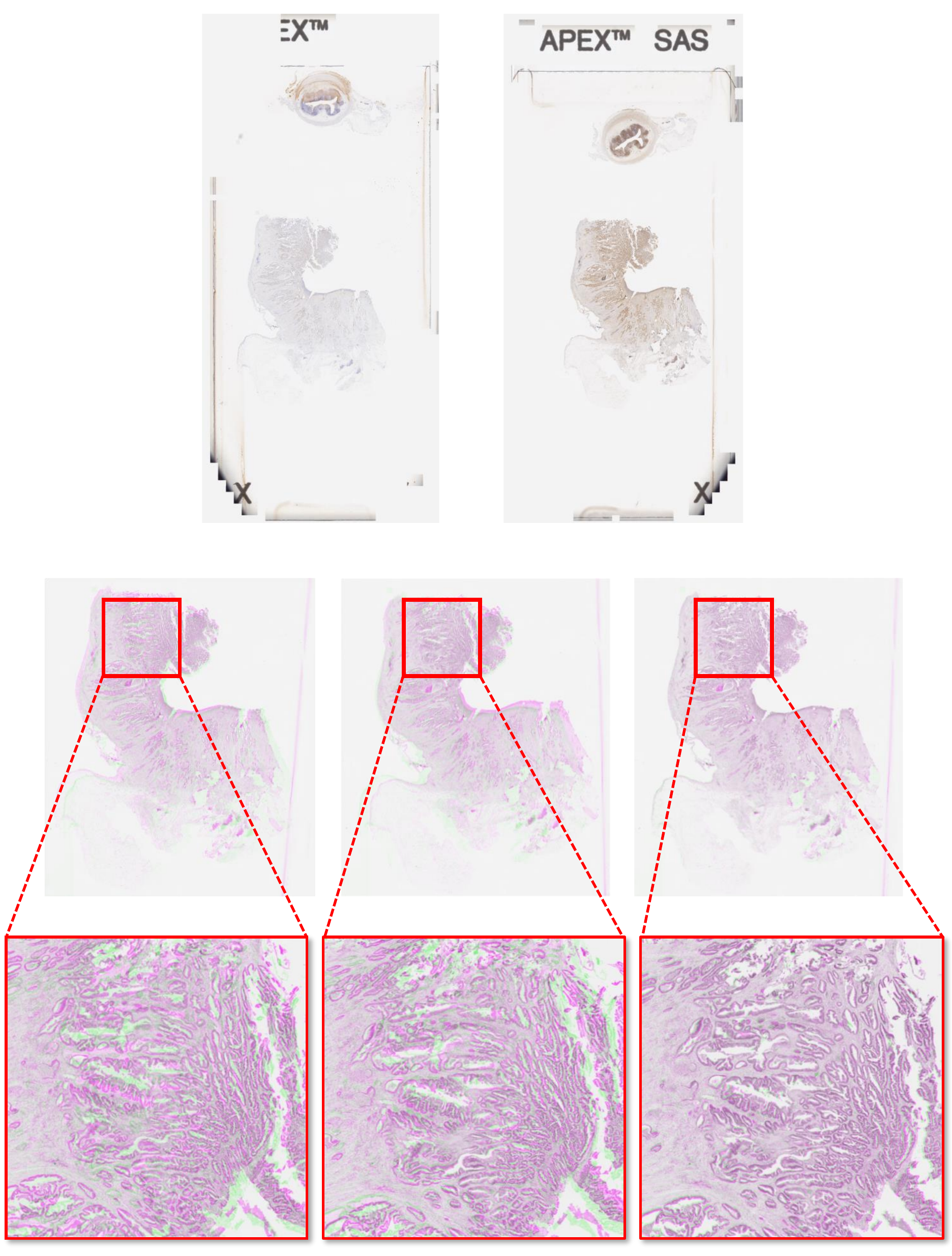}
\caption{Demonstration of the efficacy of our proposed rigid registration in a step-wise manner using overlay of reference and registered moving images. Top row shows reference and pre-aligned moving images. The middle row shows the overlay of corresponding tissue region: left, middle and right overlay images are shown as the output of the pre-alignment, the tissue based alignment and the block-wise alignment, respectively. The bottom row shows the zoomed in overlay patches. Reference and registered moving images are shown in green and purple colours, respectively. }
\label{fig:df_steps_overlaid_images}
\end{figure}

Each step in our pipeline is adopted to improve the alignment. The quantitative results in Table \ref{tab:reg_results_stepwise} demonstrate that our first three steps result in improvement of the rigid alignment between the two images while the last step has been able to tackle the non-linear deformation. Box plots shown in Figure \ref{fig:box_plot_comet} demonstrate the reduction of median rTRE after each step in our pipeline. The pre-alignment step has significantly improved the median rTRE. Figure \ref{fig:df_steps_overlaid_images} shows overlay false colour images of reference and registered moving images along with the zoomed-in visual fields. In overlay images, the extent of the green colour indicates the extent of misalignment. It can be seen that the misalignment is improved after each step.

\begin{table}
\centering
\begin{tabular}{c|c|c|c|}
\cline{2-4}
                                           & \textbf{AMrTRE} & \textbf{MMrTRE} & \textbf{AMaxrTRE} \\ \hline
\multicolumn{1}{|c|}{Initial}              &  0.1797  &  0.1068  &  0.2323   \\ \hline
\multicolumn{1}{|c|}{Pre-alignment}        &  0.0109  &  0.0087  &  0.0220   \\ \hline
\multicolumn{1}{|c|}{Tissue Transform}     &  0.0073  &  0.0046  &  0.0186    \\ \hline
\multicolumn{1}{|c|}{Block-wise Transform} &  0.0063  &  0.0041  &  0.0176    \\ \hline
\multicolumn{1}{|c|}{Non-rigid Transform}  &  0.0031  &  0.0014  &  0.0187    \\ \hline
\end{tabular}
\caption{Quantitative results generated using the COMET dataset. These results demonstrate that the average error is reduced with each alignment step in our pipeline.}
\label{tab:reg_results_stepwise}
\end{table}

\begin{figure}
\centering
\includegraphics[width=\linewidth]{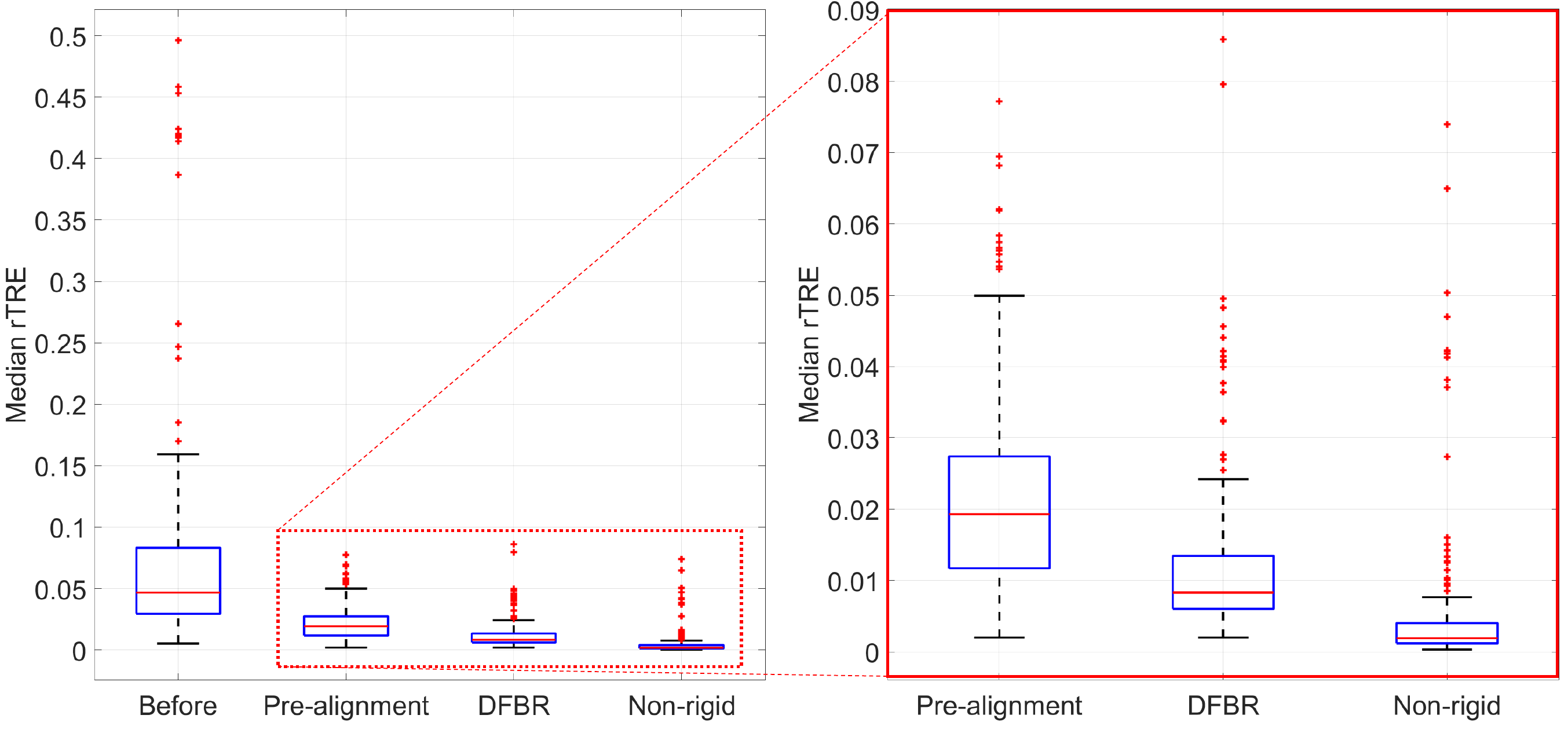}
\caption{Demonstration of the median rTRE before and after each registration module using a box plot. The median rTRE is computed for the ANHIR training set only.   }
\label{fig:box_plot_anhir}
\end{figure}

\textbf{ANHIR Dataset:}
Similar to the COMET dataset, we evaluate each step of our registration pipeline on the ANHIR dataset. As the organisers made landmarks of the training set publicly available, we first evaluate our pipeline on the training set only. The improvement in median rTRE (training set only) after each step of our method is shown in Figure \ref{fig:box_plot_anhir}. In Table \ref{tab:anhir_metric_comparison}, we show values of different metrics computed for the MEVIS and our proposed methods. Our results compare favourably in terms of aggregation average, median rTRE, maximum rTRE and robustness for the training set. We compare our results to those of the MEVIS team to show the effectiveness of the proposed method.




\begin{table}
	\resizebox{\linewidth}{!}{%
		\begin{tabular}{|c|cc|cc|cc|c|}
			\hline
			\multicolumn{1}{|c|}{\multirow{2}{*}{\textbf{Method}}} & \multicolumn{2}{c|}{\textbf{Average rTRE}}              & \multicolumn{2}{c|}{\textbf{Median rTRE}}               & \multicolumn{2}{c|}{\textbf{Max rTRE}}                  & \multicolumn{1}{c|}{\textbf{Average}}                \\ \cline{2-7} 
			\multicolumn{1}{|c|}{}                                 & \multicolumn{1}{c|}{\textbf{Average}} & \textbf{Median} & \multicolumn{1}{c|}{\textbf{Average}} & \textbf{Median} & \multicolumn{1}{c|}{\textbf{Average}} & \textbf{Median} & \multicolumn{1}{c|}{\textbf{Robustness}}  \\ \hline
			MEVIS\_Train                                           & \multicolumn{1}{c|}{0.0061}           & 0.0030          & \multicolumn{1}{c|}{0.0049}           & 0.0019          & \multicolumn{1}{c|}{0.0271}           & 0.0183          & \multicolumn{1}{c|}{0.9806}                     \\ \hline
			Ours\_Train                                            & \multicolumn{1}{c|}{0.0065}           & 0.0030          & \multicolumn{1}{c|}{0.0049}           & 0.0019          & \multicolumn{1}{c|}{0.0300}           & 0.0207          & \multicolumn{1}{c|}{0.9688}                    \\ \hline \hline
			MEVIS\_Eval                                            & \multicolumn{1}{c|}{0.0044}           & 0.0027          & \multicolumn{1}{c|}{0.0029}           & 0.0018          & \multicolumn{1}{c|}{0.0251}           & 0.0188          & \multicolumn{1}{c|}{0.9880}                      \\ \hline
			Ours\_Eval                                             & \multicolumn{1}{c|}{0.0046}           & 0.0028          & \multicolumn{1}{c|}{0.0031}           & 0.0017          & \multicolumn{1}{c|}{0.0252}           & 0.0197          & \multicolumn{1}{c|}{0.9842}                    \\ \hline \hline
			MEVIS\_All                                             & \multicolumn{1}{c|}{0.0052}           & 0.0029          & \multicolumn{1}{c|}{0.0039}           & 0.0018          & \multicolumn{1}{c|}{0.0261}           & 0.0186          & \multicolumn{1}{c|}{0.9845}                     \\ \hline
			Ours\_All                                              & \multicolumn{1}{c|}{0.0055}           & 0.0029          & \multicolumn{1}{c|}{0.0040}           & 0.0018          & \multicolumn{1}{c|}{0.0275}           & 0.0203          & \multicolumn{1}{c|}{0.9768}                      \\ \hline
	\end{tabular}}
	\caption{Quantitative results of the winning team and our pipeline on the ANHIR dataset. First row in the header represents the aggregation method for an image pair and second row represents the aggregation method for all pairs in a set. }
	\label{tab:anhir_metric_comparison}
\end{table}

\subsection{Comparative Results}

In Table \ref{tab:reg_results_comparison}, we compare our DFBR method with case-wide registration, proposed in \cite{Nick_Reg_Thesis_2017}. Trahearn \textit{et al} \cite{Nick_Reg_Thesis_2017} proposed a novel approach based on MSER features for the alignment of multi-IHC CRC sections. This approach extracts features for each MSER detected from the preprocessed H stain channel. These features are utilised for finding the best pair of MSERs and are used as corresponding control points for estimating rigid transformation. The authors employed this approach for finding the optimal order of the sections for estimating the transformation of the whole stack. We present comparative results for three different settings: 1) IHC vs IHC, 2) H\&E vs IHC and 3) the combination of the first two. Typically, tissue sections of around 3-5 microns in thickness are sliced from the tissue block. The exact thickness of the tissue section could not be retrieved for the COMET sections. However, if we consider it to be 5 microns then the spatial distance between the H\&E and IHC slides would range between 30-50 microns whereas for IHC images it would be 5 to 15 microns, making the registration of the H\&E and IHC slides challenging not just because of staining differences but also due to the morphological differences between them. Our DFBR approach outperforms the case-wide approach \cite{Nick_Reg_Thesis_2017} in both settings. Most importantly, the results demonstrate that our DFBR approach can align images even when the tissue structures vary significantly between two images.


In terms of processing time, the DFBR approach is comparable to the MSER feature based approach on registering pair of images rather than the whole stack. For any image pair, the DFBR approach takes a consistent amount of time with a mean processing time of 14 seconds within 1 standard deviation across 105 pairs of images. The consistency is because there are a fixed number of feature points against which feature matching is performed. Whereas, the processing time for the pairwise MSER approach is dependent on the number of MSERs detected. In our experiments with 105 pairs of images, it took 17$\pm$13 seconds. Similarly, processing time varies for case-wide registration; on average it takes around 5 minutes for transforming the whole stack of images \cite{Nick_Reg_Thesis_2017}.

\begin{table}
\resizebox{\linewidth}{!}{%
\begin{tabular}{ll|c|c|c|c|}
\cline{3-6}
                                                     &           & \textbf{Avg(Med(rTRE))} & \textbf{Med(Med(rTRE))} & \textbf{Avg(Max(rTRE))} & \textbf{Robustness} \\ \hline
\multicolumn{1}{|l|}{\multirow{3}{*}{IHC vs IHC}} &  Case-wide & 0.0053                  & 0.0033                  & 0.0114                  & 0.9801              \\
\cline{2-6}
\multicolumn{1}{|l|}{}                               & DFBR & \textbf{0.0039}                  & \textbf{0.0028}                  & \textbf{0.0091}                  & \textbf{0.9948}              \\
\hline \hline

\multicolumn{1}{|l|}{\multirow{3}{*}{H\&E vs IHC}} &  Case-wide & 0.0120         & \textbf{0.0073}                  & 0.0381                  & 0.9529   \\
\cline{2-6}
\multicolumn{1}{|l|}{}                               & DFBR & \textbf{0.0111}         & 0.0102                  & \textbf{0.0344}                  & \textbf{1}   \\
\hline \hline

\multicolumn{1}{|l|}{\multirow{3}{*}{All}}   &  Case-wide & 0.0075                  & 0.0049                  & 0.0200                  & 0.9715              \\
\cline{2-6}
\multicolumn{1}{|l|}{}                               & DFBR & \textbf{0.0063}                  & \textbf{0.0041}                  & \textbf{0.0176}                  & \textbf{0.9965}              \\
\hline
\end{tabular}}
\caption{Comparative results of registration using a case-wide \cite{Nick_Reg_Thesis_2017} and DFBR approaches. This comparative analysis is presented for two different sets: IHC vs IHC and H\&E vs IHC. }
\label{tab:reg_results_comparison}
\end{table}

\section{Summary}
In this paper, we presented a deep feature matching approach which is shown to outperform the hand-crafted feature based approach. In the DFBR framework, we introduced a pre-alignment step which produces a roughly aligned image pair. Tissue segmentation is an essential pre-processing step to exclude feature points from the texture sparse region. Our experiments with deep features showed that a good alignment is difficult to obtain in situations when the slide has more fatty tissue. Therefore, a good tissue segmentation which considers the fatty region as a background is required.  

In a digital stack of tissue sections, it is likely that we will observe differences in the tissue structures across the whole stack. This is due to the thickness of each sections in relation to the size of the tissue structures. Smaller structures like cells are likely to be seen between the consecutive sections. However as the distance between the two sections increases, same cells and even some bigger tissue structures (such as glands) will not persist. A registration method should be able to perform alignment in the presence of architectural differences between the two sections. Our deep feature method has been shown to outperform an existing MSER based case-wide method \cite{Nick_Reg_Thesis_2017} when the distance between image pairs is around 50 microns. 

\bibliographystyle{ieeetr}

\end{document}